\documentclass[12pt]{article}
\usepackage{amsmath}
\begin{document}
\begin{flushright}
KANAZAWA-10-13\\
December, 2010
\end{flushright}
\vspace*{1cm}

\begin{center}
{\Large\bf Signals of dark matter in a supersymmetric two dark matter model}
\vspace*{1cm}

{\Large Hiroki Fukuoka}\footnote{e-mail:~fukuoka@hep.s.kanazawa-u.ac.jp},
{\Large Daijiro Suematsu}\footnote{e-mail:~suematsu@hep.s.kanazawa-u.ac.jp}
{\Large and Takashi Toma}\footnote{e-mail:~t-toma@hep.s.kanazawa-u.ac.jp}
\vspace*{1cm}

{\it Institute for Theoretical Physics, Kanazawa University, 
\\ Kanazawa 920-1192, Japan}
\end{center}
\vspace*{1.5cm} 

\noindent
{\Large\bf Abstract}\\
Supersymmetric radiative neutrino mass models have often two dark matter
candidates. One is the usual lightest neutralino with odd $R$ parity 
and the other is a new neutral particle whose stability is guaranteed 
by a discrete symmetry that forbids tree-level neutrino Yukawa couplings. 
If their relic abundance is comparable, dark matter phenomenology can be
largely different from the minimal supersymmetric standard model (MSSM).  
We study this in a supersymmetric radiative neutrino mass model with 
the conserved $R$ parity and a $Z_2$ symmetry weakly broken by the
anomaly effect. The second dark matter with odd parity of this new $Z_2$ 
is metastable and decays to the neutralino dark matter. 
Charged particles and photons associated to this decay can cause the 
deviation from the expected background of the cosmic rays.
Direct search of the neutralino dark matter is also expected to show 
different features from the MSSM since the relic abundance 
is not composed of the neutralino dark matter only. 
We discuss the nature of dark matter in this model by analyzing these
signals quantitatively.

\newpage
\section{Introduction}
The explanation of small neutrino masses \cite{oscil} and 
dark matter \cite{wmap} seems to be a key ingredient to consider 
physics beyond the standard model (SM). 
An interesting possibility of such extensions may be models which can
closely relate neutrino masses to dark matter (DM).
In this kind of models, a discrete symmetry is often introduced 
to forbid tree level Dirac neutrino masses. 
Some of additional particles introduced to commit 
neutrino mass generation have its charge such that it can forbid 
the lightest one to decay into the SM particles. 
This stable particle becomes DM. 
This DM is a crucial ingredient of the neutrino mass generation 
in this scenario.

The radiative seesaw model proposed by Ma \cite{Ma:2006km} is its simple
and interesting example.\footnote{A lot of radiative neutrino mass models
exist now. Phenomenology including the DM nature in such models 
has been studied in a lot of works \cite{cdmmeg,lflavor,ext,scdm,fcdm,ncdm}.}
Both the numbers of new particles and free parameters are comparably small.
Its supersymmetric extension is also 
straightforward \cite{sma,fks}.\footnote{A relevant supersymmetric model
is also considered in a different context in \cite{e6}.}
Moreover, if we introduce an anomalous U(1) symmetry 
in this extension \cite{anomu},
we could explain the origin of the discrete symmetry, required hierarchical
structure of both couplings and masses due to the Frogatt-Nielsen
mechanism \cite{fn,yh}.  
Both neutrino oscillation data and DM relic abundance can also be 
explained consistently with lepton flavor violating processes such
as $\mu\rightarrow e\gamma$.
A characteristic feature in such an extension with $R$ parity
conservation is that the model has two DM
candidates\footnote{Multicomponent DM and its phenomenology are studied
in a different model \cite{flnp}.}.
One is the lightest superparticle whose stability is guaranteed by the
$R$ parity.  The other one is a new particle introduced for the neutrino
mass generation and its stability is guaranteed by the new $Z_2$ symmetry. 
As a result, the model shows discriminative differences from the
ordinary minimal supersymmetric SM (MSSM) in the DM search. 
For example, if the recently reported cosmic ray anomalies 
\cite{pamela,fermi} are considered as the DM signature of 
the model, they may be explained not by the DM 
annihilation \cite{mindep,sommerfeld,bwenhance}
as in the MSSM but by the DM decay 
\cite{it,decay,gravitino,rparity,hidden}.
In fact, if the $Z_2$ symmetry is violated by the anomaly effect,
the DM guaranteed its stability by the $Z_2$ symmetry can decay into 
the lightest neutralino \cite{fks,anomu}.
Direct search of the DM could also show the difference from the MSSM.

In this paper, we study signals of the DM in the supersymmetric
extension of the Ma model.
The model is considered as an effective model due to spontaneously
broken anomalous U(1) gauge symmetry.
It naturally brings the weakly broken $Z_2$ symmetry to the model 
in addition to the conserved $R$ parity. 
We discuss signatures due to the decay of the unstable DM and also 
the direct detection of the DM through the elastic
scattering with nuclei.

The paper is organized as follows. In section 2 we address the model and
explain the nature of the DM sector which is imposed 
by various experimental results.
In section 3 several signals expected in the DM sector are analyzed.
In particular, the decaying DM is studied to explain the cosmic ray 
anomalies reported recently.
A feature of the monochromatic gamma yielded through the DM radiative 
decay is also studied. Finally, we discuss the direct search of the DM.
Section 4 is devoted to the summary.
  
\section{A supersymmetric radiative neutrino mass model}
The radiative seesaw model proposed in \cite{Ma:2006km} is an 
extension of the SM with three right-handed neutrinos 
and an inert doublet scalar. 
The latter is assumed to have no vacuum expectation value (VEV) 
and no coupling with quarks.
Although the model is very simple and has several interesting features 
\cite{cdmmeg,lflavor,ext}, it has some faults, that is, the existence of an
extremely small coupling and the ordinary hierarchy problem.
These may be improved by extending the model with supersymmetry and an
anomalous U(1) symmetry \cite{anomu}.
We focus our present study on this model, which has a $Z_2$ symmetry 
as a remnant subgroup after the spontaneous symmetry breaking of 
this anomalous U(1). 
Matter contents of the model and their $Z_2$ charge are 
summarized in Table 1. 

\begin{figure}[t]
\begin{center}
\begin{tabular}{|c|ccccccccccc|} \hline
$\Psi_\alpha$ &$Q_i$ & $U^c_i$ & $D^c_i$ & $L_i$ & $E^c_i$ &$H_u$&$H_d$ 
& $N^c_i $ & $\eta_u $ & $\eta_d$ & $\phi $  \\ \hline
$R$  & $-$ & $-$ &$-$& $-$ & $-$ & $+$&  $+$ &  $+$ &  $-$ 
& $-$& $-$ \\ \hline
$Z_2$ & $+$ & $+$ &$+$& $+$ & $+$ &  $+$&  $+$& $-$ &  $-$ 
& $-$& $-$  \\ \hline
\end{tabular}
\end{center}
\vspace*{3mm}

{\footnotesize{\bf Table 1}~~Matter contents and their quantum number. 
$Z_2$ is a remnant symmetry of the assumed anomalous U(1) caused by the
symmetry breaking at a high energy region.}
\end{figure}

The most general superpotential invariant under the imposed symmetry is
\begin{eqnarray}
W&=&h_{ij}^U Q_{i} U_{j}^c  H_u
+ h_{ij}^DQ_{i} D_{j}^c H_d
+ h_{i}^EL_{i} E_{i}^c H_d
+\mu_H H_u H_d \nonumber \\
&+&h_{ij}^NL_{i}N_{j}^c \eta_u
+\lambda_u\eta_u H_d \phi
+\lambda_d\eta_d H_u \phi
+\mu_\eta\eta_u\eta_d
+ \frac{1}{2}M_iN_i^cN_i^c 
+\frac{1}{2}\mu_\phi\phi^2 \nonumber \\
&+&c_iM_{\rm pl}e^{-b_i}L_i\eta_u.
\label{superpot}
\end{eqnarray}
This can be obtained as the low energy effective theory
through the spontaneous breaking of the 
anomalous U(1) as shown in \cite{anomu}. 
The last term in $W$ is induced by an anomaly effect \cite{gs,adm}.
This term breaks the $Z_2$ symmetry very weakly if $b_i$ is large enough.
Since the $Z_2$ symmetry is not exactly conserved, the lightest field with
odd parity of the $Z_2$ is unstable. However, the lifetime can be longer
than the age of universe and it behaves as the DM. 
Thus, we have two DM components in the model as long as the $R$ parity 
is conserved.

Soft supersymmetry breaking terms associated with 
the superpotential $W$ are introduced as follows,
\begin{eqnarray}
{\cal L}_{SB}
&=&-\tilde m_{\eta_u}^2\tilde{\eta}_u^\dag\tilde{\eta}_u-
\tilde m_{\eta_d}^2\tilde{\eta}_d^\dag\tilde{\eta}_d
-\tilde m_{N^c}^2 \tilde{N^c}^\dag\tilde N^c
-\tilde m_{\phi}^2 \tilde{\phi}^\dag\tilde\phi \nonumber \\
&&
+A(h_{ij}^N\tilde{L}_i\tilde{N}^c_j\tilde{\eta}_u
+\lambda_u \tilde{\eta}_u {H}_d\tilde\phi+
 \lambda_d \tilde{\eta}_d{H}_u\tilde\phi + {\rm h.c.}) \nonumber \\
&&-B\left(\mu_\eta\tilde{\eta}_u \tilde{\eta}_d+ 
\frac{1}{2}\mu_\phi\tilde{\phi}^2+\frac{1}{2}M_i \tilde N_i^{c2}
+c_iM_{\rm pl}e^{-b_i}\tilde{L}_i\tilde{\eta}_u 
+{\rm h.c.}\right).
\label{softsb}
\end{eqnarray}
The scalar components are represented by putting a tilde on the 
character of the corresponding chiral superfield except
for the ordinary Higgs chiral superfields $H_u$ and $H_d$. 
Universality of soft supersymmetry breaking $A$ and $B$ parameters 
is assumed, for simplicity. 
Moreover, we confine our following consideration to
the case where soft masses for all the scalar partners 
are flavor diagonal and universal unless we mention it. 
They are denoted by $m_0$.

Neutrino masses are generated through the one-loop diagram as discussed
in \cite{anomu}.
If we focus our attention to the special flavor structure 
for neutrino Yukawa couplings such as \cite{lflavor}
\begin{equation}
h_{ei}^N=0, \quad h_{\mu i}^N=h_{\tau i}^N\equiv
 |h_i|e^{i\varphi_i}~~(i=1,2), \qquad
h_{e3}^N=h_{\mu 3}^N=-h_{\tau 3}^N\equiv |h_3|e^{i\varphi_3},
\label{yukawa}
\end{equation}
the neutrino mass matrix is found to be expressed as
\begin{equation}
{\cal M}_\nu=\left(
\begin{array}{ccc}
0 & 0 & 0\\ 0 & 1 & 1 \\ 0 & 1 & 1 \\ \end{array}\right)
(h_{\tau 1}^2\Lambda_1+h_{\tau 2}^2\Lambda_2)
+\left(
\begin{array}{ccc}
1 & 1 & -1\\ 1 & 1 & -1 \\ -1 & -1 & 1 \\ \end{array}\right)
h_{\tau 3}^2\Lambda_3.
\label{nmass}
\end{equation}
This mass matrix induces the tri-bimaximal MNS 
matrix.\footnote{The charged lepton mass matrix is assumed to be diagonal 
when we consider the flavor structure of neutrino Yukawa 
couplings (\ref{yukawa}).}
Mass scales for the neutrino masses are determined by $\Lambda_i$,
which is defined as
\begin{eqnarray}
&&\Lambda_i=\frac{\bar\lambda v^2 M_i}
{32\pi^2}\Big(g(M_i,m_{\eta +})-g(M_i,m_{\eta -})\Big), \nonumber \\
&&g(m_a,m_b)=\frac{m_a^2 -m_b^2+m_a^2\ln(m_b^2/m_a^2)}{(m_a^2-m_b^2)^2},
\qquad \bar\lambda=\frac{\lambda_u\lambda_d\tan\beta}{1+\tan^2\beta},
\label{mscale}
\end{eqnarray}
where $\langle H_u^0\rangle=v\sin\beta$ and $\langle H_d^0\rangle=v\cos\beta$.
$\lambda_{u,d}$ are assumed to be real, for simplicity. 
$m_{\eta\pm}^2$ are the mass eigenvalues of the neutral scalar 
components of $\eta_{u,d}$, which are defined as 
$m_{\eta\pm}^2\simeq \mu_\eta^2+m_0^2\pm B\mu_\eta$.
If $M_i$ and $m_{\eta\pm}$ have the values of $O(1)$~TeV,
mass eigenvalues of neutrinos can be suitable values as long as 
$\lambda_u$ and $\lambda_d$ take very small values such as 
$\lambda_u\lambda_d=O(10^{-8})$. 

Before proceeding the analysis of the DM phenomenology, 
it is useful to address free parameters in the neutrino sector of the model. 
The relevant parameters are summarized as 
$\bar\lambda$, $\mu_\eta$, $m_0$, $B$ and also
$|h_i|$, $\varphi_i$, $M_i~(i=1,2,3)$. 
We restrict our study to the case with $M_1~{^<_\sim}~M_2<M_3$ which
allows the coannihilation of $\psi_{N_1}$ and
$\psi_{N_2}$ (the fermionic components of $N_1^c$ and $N_2^c$).
 We consider this case since it brings
an interesting aspect in DM phenomenology as seen below.
Since one eigenvalue of (\ref{nmass}) is zero,
the neutrino oscillation data tell us that remaining eigenvalues should be 
$\sqrt{\Delta m^2_{\rm atm}}$ and $\sqrt{\Delta m_{\rm sol}^2}$. 
This imposes the parameters to satisfy the relations 
\begin{equation}
|h_1^2+h_2^2|\Lambda_1\simeq \frac{\sqrt{\Delta m_{\rm atm}^2}}{2}, \quad
|h_3|^2\Lambda_3\simeq \frac{\sqrt{\Delta m_{\rm sol}^2}}{3}.
\label{c-oscil}
\end{equation}
Thus, after using these relations, the free parameters in the neutrino
sector can be confined to
\begin{equation}
\bar\lambda, \quad M_1, \quad M_3, \quad \mu_\eta, 
\quad m_0, \quad B, \quad \varphi_i.
\label{para0}
\end{equation}

\input epsf
\begin{figure}[t]
\begin{center}
\epsfxsize=6cm
\leavevmode
\epsfbox{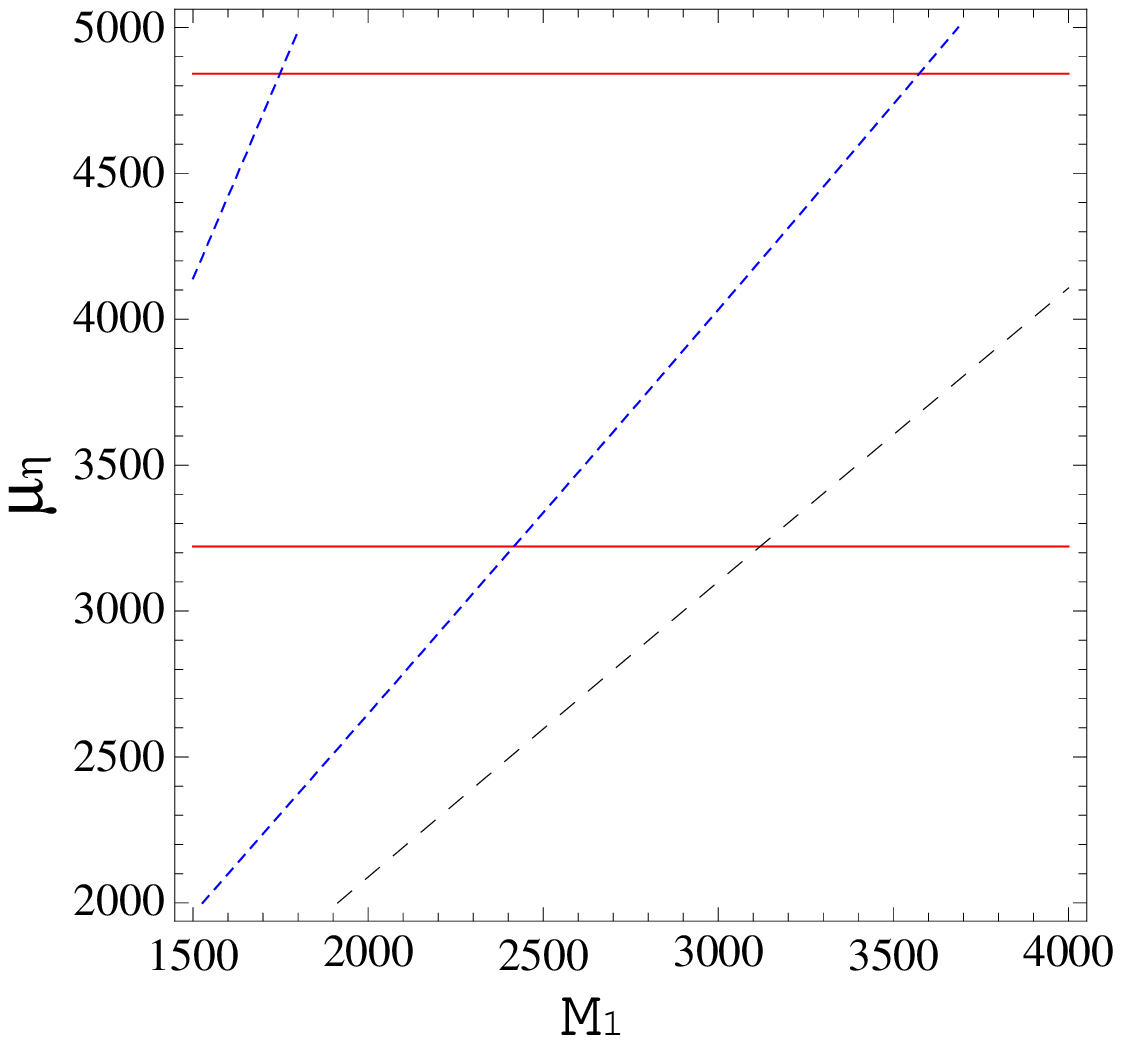}
\hspace*{5mm}
\epsfxsize=7cm
\leavevmode
\epsfbox{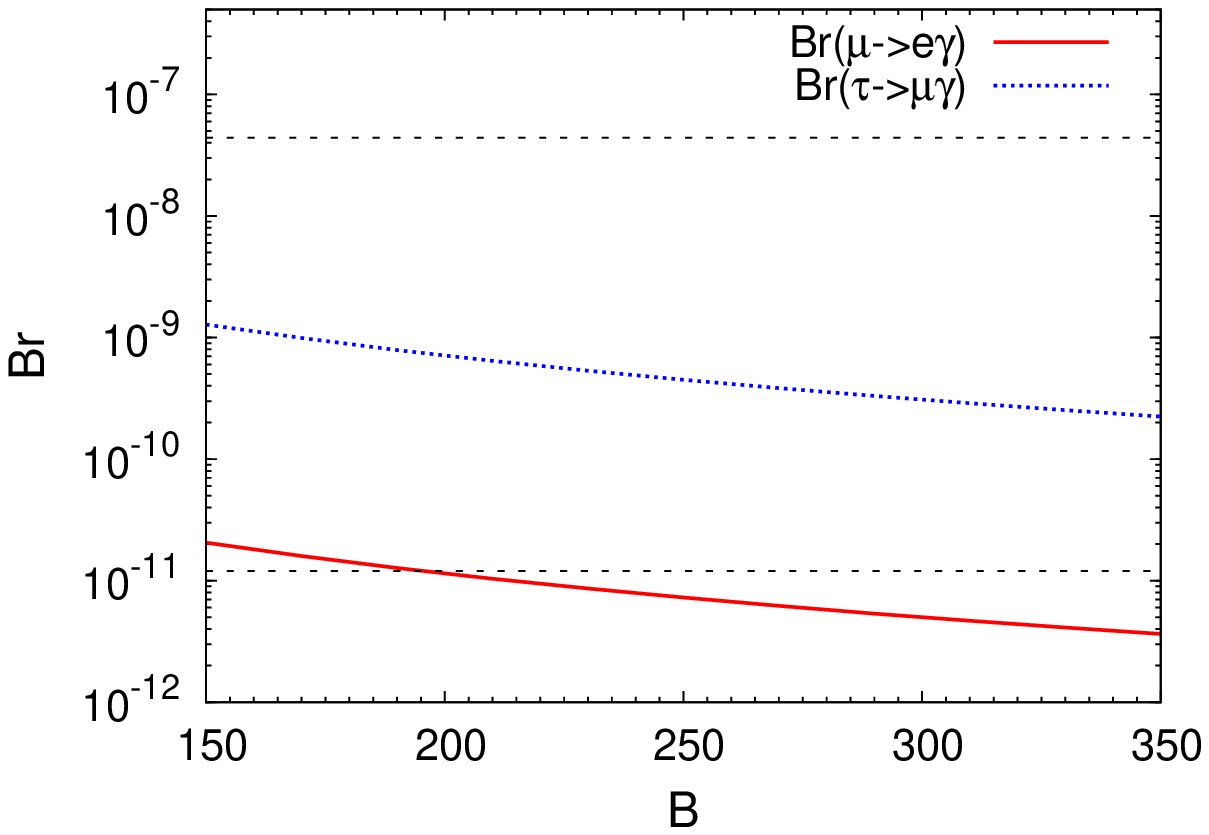}
\end{center}

\vspace*{-3mm}
{\footnotesize {\bf Fig.~1}~~The left frame shows a parameter region 
in the $(M_1,\mu_\eta)$ plane which is consistent with the 
neutrino oscillation data and the LFV constraints for $B=250$~GeV. 
An upper red solid line represent a contour for 
${\rm Br}(\mu\rightarrow e\gamma)=1.2\times 10^{-11}$ which is the
 present experimental upper bound \cite{meg} and another red solid line 
represents a reference value ${\rm Br}(\mu\rightarrow e\gamma)=
0.6\times 10^{-11}$. Blue dotted lines represent the contours 
of ${\rm Br}(\tau\rightarrow \mu\gamma)$ for reference values $2\times
 10^{-9}$  (the left one) and $0.6\times 10^{-9}$ (the right one). Its
experimental upper bound is $4.4\times 10^{-8}$ \cite{babar}. 
$M_1=m_{\eta-}$ is satisfied on a black dashed line. 
In the right frame these branching ratios are plotted as a function of
 $B$ for $M_1=3200$~GeV and $\mu_\eta=3600$~GeV.
It shows an allowed region of $B$ which is consistent 
with the neutrino oscillation data and the LFV constraints. 
A red solid line and a blue dotted line represent
${\rm Br}(\mu\rightarrow e\gamma)$ and ${\rm Br}(\tau\rightarrow
 \mu\gamma)$ in this model, respectively. 
Thin black dotted lines represent the experimental bounds 
for ${\rm Br}(\mu\rightarrow e\gamma)$ (the lower one)  and 
${\rm Br}(\tau\rightarrow\mu\gamma)$ (the upper one).}
\end{figure}

Here we search parameter regions consistent with the experimental data for
the lepton sector. For this purpose, we can use neutrino oscillation 
data \cite{oscil} and the constraints from lepton flavor violating 
processes (LFV) such as 
${\rm Br}(\mu\rightarrow e\gamma)<1.2\times 10^{-11}$ \cite{meg} and
 ${\rm Br}(\tau\rightarrow\mu\gamma)<4.4\times 10^{-8}$ \cite{babar}.
We fix a part of the parameters listed in eq.~(\ref{para0}) as\footnote{
In this analysis we use the parameters different from the ones 
used in \cite{anomu}. It could cause some differences for the bounds of
parameters between these two cases. For example, since we use a larger 
value of $M_3$ than the one in \cite{anomu} here, the LFV constraints are
satisfied for a smaller $m_0$ value compared with the one discussed 
there.  These parameters are adopted here since they are favorable 
for the explanation of the cosmic ray anomalies as discussed later.} 
\begin{equation}
\bar\lambda=1.16\times 10^{-9}, \quad  m_0=400~{\rm GeV}, \quad
M_3=9000~{\rm GeV}, \quad \varphi_1-\varphi_2=0.
\label{para}
\end{equation}
If we use these parameters in eq.~(\ref{c-oscil}) and the formulas for
the branching ratio of the LFV \cite{anomu}, 
we can find a parameter region in the $(M_1,\mu_\eta)$ plane 
which is consistent with both the neutrino oscillation
data and the constraints from the LFV.
It is plotted in the left frame of Fig.~1 for $B=250$~GeV.  
In this figure, a region sandwiched by the upper red solid line which
represents the contour ${\rm Br}(\mu\rightarrow e\gamma)=1.2\times 10^{-11}$ 
and the black dashed line which represents $M_1=m_{\eta_-}$
is an allowed region if $\psi_{N_1}$ is assumed to be the lightest 
$Z_2$ odd particle. 
It shows that the LFV constraints can be satisfied for 
$\mu_\eta~{^<_\sim}~4850$~GeV. This result does not depend on the $m_0$
value sensitively in the region where $M_1,\mu_\eta \gg m_0$ is
satisfied as long as $B$ is fixed in the region $B\ll M_1,~\mu_\eta$.
If we take a larger $\bar\lambda$, the condition (\ref{c-oscil}) can be
satisfied by smaller neutrino Yukawa couplings. 
In that case we note that the LFV constraints become weaker.
It is interesting that this allowed region in the $(M_1,\mu_\eta)$ 
plane relevant to the following analysis is within the reach 
of $\mu\rightarrow e\gamma$ search in the MEG experiment.
It aims to search for $\mu^+\rightarrow e^+\gamma$ decay with a
sensitivity of a few $\times 10^{-13}$ \cite{meg1}.
 
In the right frame of Fig.~1, we plot the branching ratio for
$\mu\rightarrow e\gamma$ and $\tau\rightarrow\mu\gamma$
as a function of $B$ for $M_1=3200$~GeV and
$\mu_\eta=3600$~GeV which are contained in the allowed region as shown 
in the left frame. This figure shows that the LFV bounds can be
satisfied for $B~{^>_\sim}~200$~GeV.
If we fix $B$ to 250~{\rm GeV} for example,
Yukawa couplings are found to have rather large values such as 
$|h_1^2+h_2^2|^{1/2}\simeq 2.98$ and $|h_3|\simeq 1.14$.
Although the values of Yukawa couplings gradually decreases for larger
values of $B$, they are always large in this figure.
These large Yukawa couplings are required to reduce the relic abundance
of $\psi_{N_1}$ with such a large mass sufficiently.\footnote{If we
assume smaller values for $M_1$ and $\mu_\eta$, small neutrino Yukawa couplings
can explain the $\psi_{N_1}$ relic abundance consistently with other
constraints by fixing $\bar\lambda$ to a larger value.}  
Although they could cause a problem for perturbativity
of the model, we can escape this fault of the model by considering
the phases of neutrino Yukawa couplings. This point is discussed below.

\section{Signals of the DM}
\subsection{Relic abundance of two DM}
The model has two types of DM candidate.
One of them is the lightest neutralino $\chi$ whose stability is 
guaranteed by the $R$ parity as in the case of the MSSM.  
The other one is the lightest neutral state
composed of the components of 
$Z_2$ odd chiral supermultiplets $N_i^c$, $\eta_{u,d}^0$ and $\phi$.
In the following study, we assume that $\psi_{N_1}$ 
(the fermionic component of $N_1^c$) is the lightest 
one among these candidates. 
Since this $Z_2$ is not an exact symmetry but is weakly broken by the
last term of $W$ through anomaly effect, $\psi_{N_1}$ is not stable. 
However, it could have a longer lifetime than the age 
of the universe as long as $b_i$ is large enough.
If this is the case, the DM relic abundance suggested by the WMAP \cite{wmap}
should be satisfied by both of these contributions.
This condition is expressed as
\begin{equation}
\Omega_{\chi}h^2+\Omega_{\psi_{N_1}}h^2=0.11.
\label{wmap}
\end{equation}
If $\Omega_\chi \gg \Omega_{\psi_{N_1}}$ is satisfied, DM physics 
is the same as the one of the MSSM. 
However, we are interested in a
different situation from the MSSM, where both of them cause the same 
order contributions.
In order to study DM physics for such a case, we search a
parameter region which brings this situation within the parameter space
discussed in the previous part.

\begin{figure}[t]
\begin{center}
\epsfxsize=7cm
\leavevmode
\epsfbox{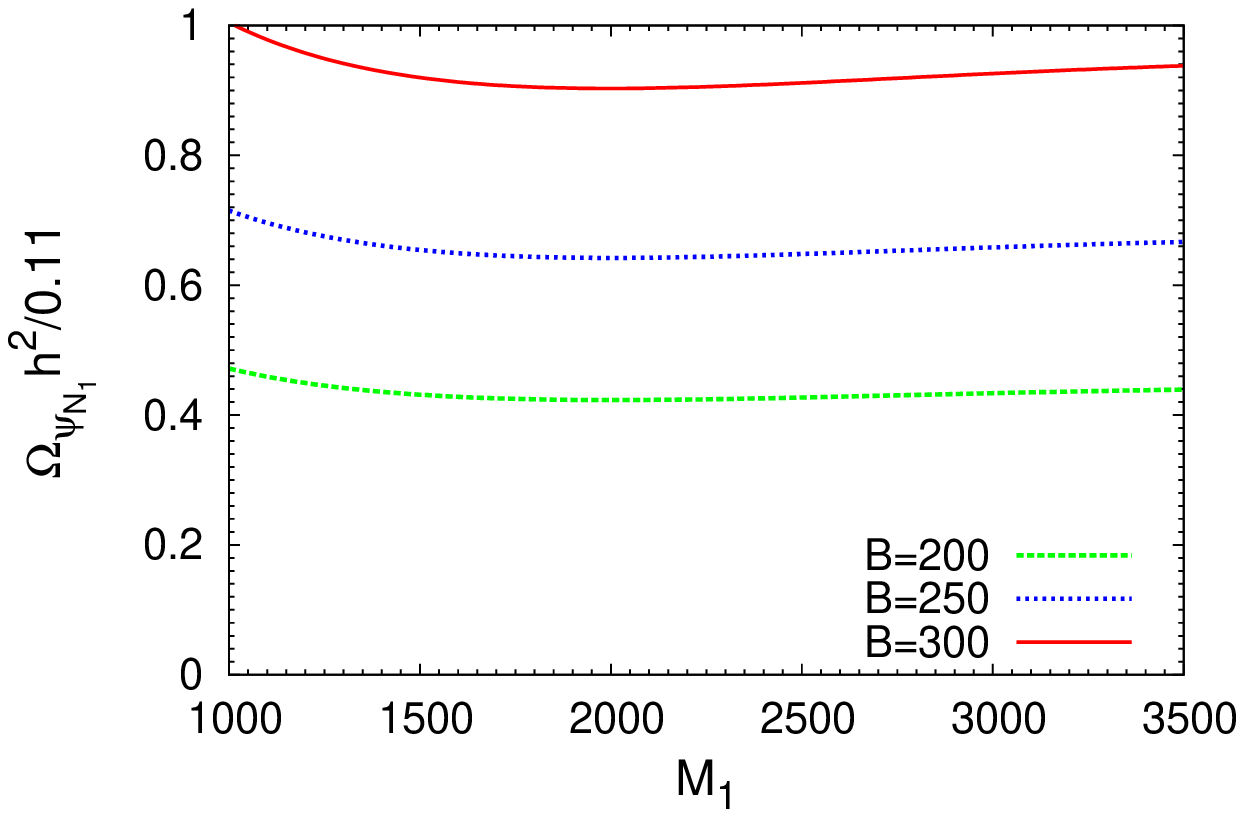}
\hspace*{5mm}
\epsfxsize=7cm
\leavevmode
\epsfbox{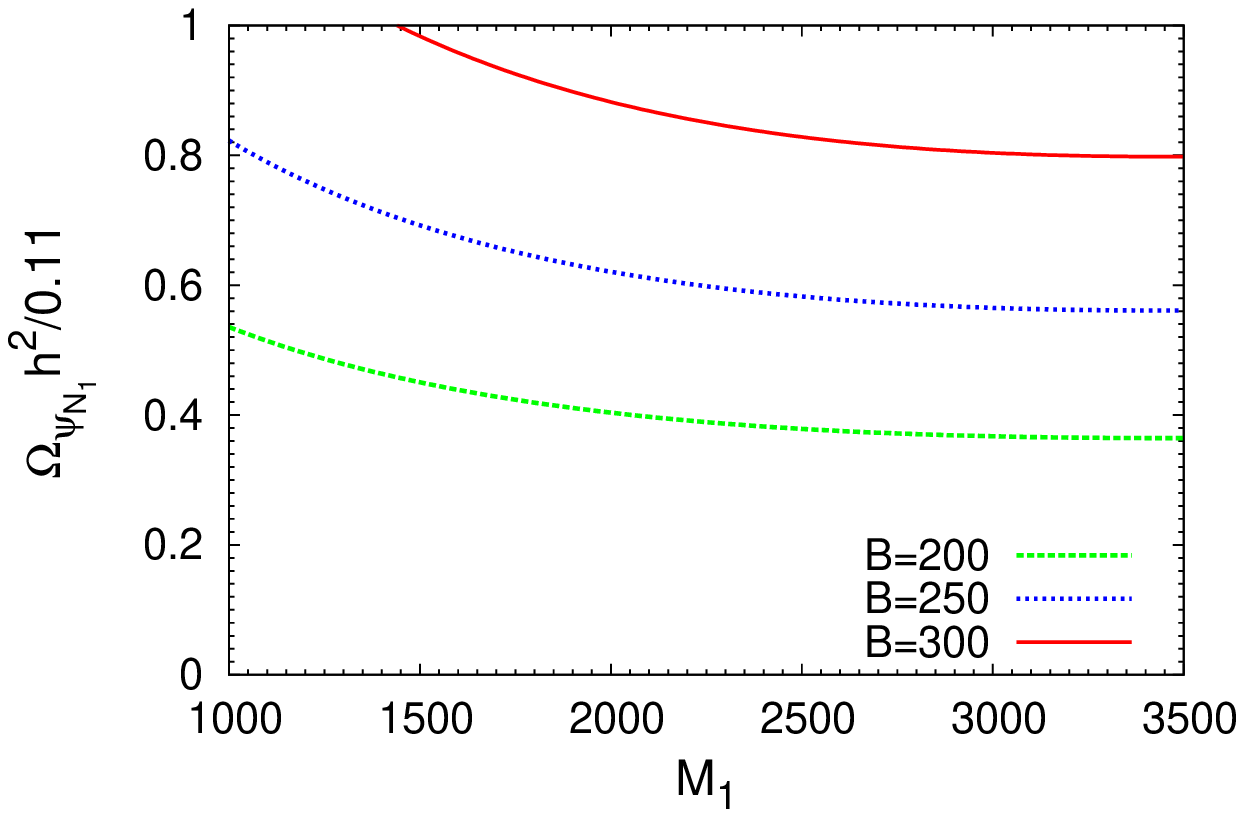}
\end{center}
\vspace*{-3mm}
{\footnotesize {\bf Fig.~2}~~The relic abundance
 $\Omega_{\psi_{N_1}}h^2/0.11$ as a function of the mass of
 $\psi_{N_1}$ for the cases $B=200,250,300$~GeV. 
$\varphi_1-\varphi_2$ is fixed to 0 (the left frame) and
 $\frac{\pi}{4}$ (the right frame), respectively. 
$\bar\lambda$ are fixed to $1.16\times 10^{-9}$ and $7.74\times 10^{-9}$
in the left and right frame, respectively. Other parameters used to 
draw these figures are explained in the text.}
\end{figure}

First, we consider the annihilation processes which determine the relic
abundance of $\psi_{N_1}$. 
The annihilation is induced through the 
$t$- and $u$-channel $\eta_u$ exchange. 
If $\psi_{N_2}$ has the almost degenerate mass 
 with $\psi_{N_1}$, we need take account of 
the coannihilation effect \cite{coann}. We suppose such a situation here.
The possible final states of such processes are composed of 
a pair of lepton and antilepton or a pair of slepton and antislepton.
Applying the method developed in \cite{coann,griest} to this model, 
we can estimate the relic abundance $\Omega_{\psi_{N_1}}h^2$.
The details can be found in \cite{anomu}.
In this estimation, we use the parameters given 
in eq.~(\ref{para}) and $\mu_\eta=3600$~GeV
which can be consistent with the neutrino oscillation data and the LFV
constraints as seen before.\footnote{
Since the difference between $M_1$ and $m_{\eta_-}$
is $10\%$ in case of $M_1=3200$~GeV, $m_0=400$~GeV and $B=200$~GeV for example,
the coannihilation of $\psi_{N_1}$ and $\eta_-$ could play some role 
for a larger $B$ \cite{coann}. However, we neglect their
coannihilation effect in this analysis.}

In the left frame of Fig.~2,
the relic abundance $\Omega_{\psi_{N_1}}h^2$
are plotted as a function of the $\psi_{N_1}$ mass $M_1$ for 
typical values of $B$. 
In the right frame we also plot the same figure for
$\bar\lambda=7.74\times 10^{-9}$ in the case of
$\varphi_1-\varphi_2=\frac{\pi}{4}$. Other parameters are fixed to the
same values as the ones in the left frame.
As noted above, since $\bar\lambda$ is fixed to the larger value
compared with the one used in the left frame, neutrino Yukawa couplings can
take smaller values keeping the consistency with the 
condition (\ref{c-oscil}).
On the other hand, $\varphi_1\not=\varphi_2$ generates 
the $s$-wave contribution in the coannihilation cross section for 
$\psi_{N_1}$ and $\psi_{N_2}$ \cite{anomu}. 
As a result, neutrino Yukawa couplings such as $|h_1^2+h_2^2|^{1/2}\simeq 1.19$ 
and $|h_3|\simeq 0.45$ are sufficient to realize the suitable relic abundance 
in case of $M_1=3200$~GeV, for example.
These values of Yukawa couplings are much smaller than 
the ones in the $\varphi_1=\varphi_2$ case. 
Because of this feature, the present LFV limits give no constraints 
on the model in this case. 
This situation is largely different from the $\varphi_1=\varphi_2$ 
case where the LFV constraints could play a crucial role to restrict the
parameter space as shown in Fig.~1. 

\begin{figure}[t]
\begin{center}
\epsfxsize=7.5cm
\leavevmode
\epsfbox{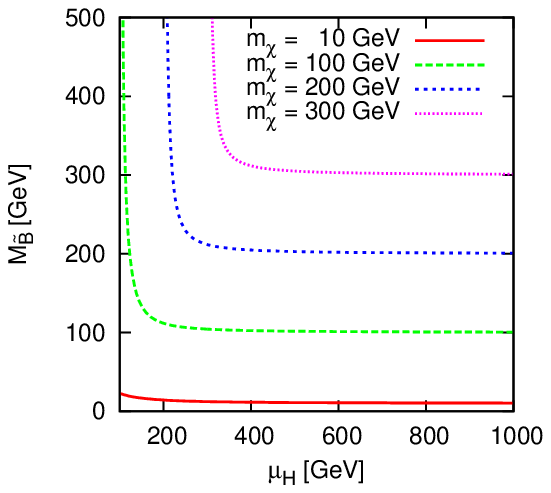}
\hspace*{1mm}
\epsfxsize=8cm
\leavevmode
\epsfbox{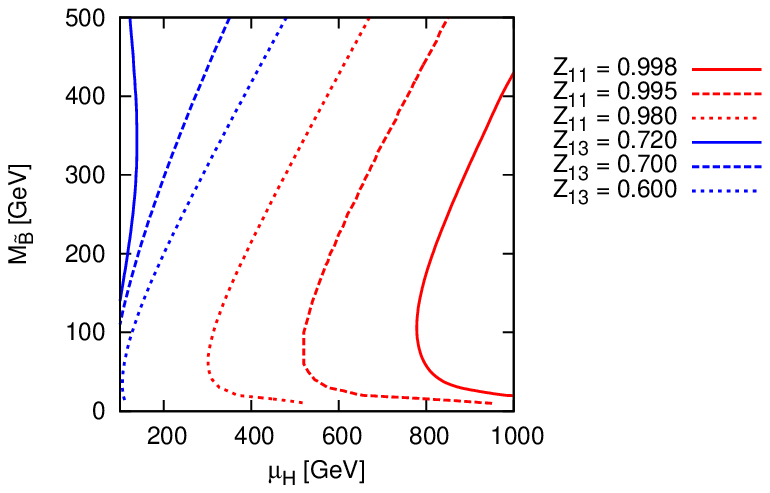}
\end{center}
\vspace*{-3mm}
{\footnotesize {\bf Fig.~3}~~Contours of the mass of the lightest neutralino
 $\chi$ (left frame) and contours of its composition $Z_{1i}$ 
(right frame) in the $(\mu_H, M_{\tilde B})$ plane. Only a region with
 $\mu_H>500$~GeV is allowed for $m_0=400$~GeV and $B>200$~GeV. }
\end{figure}

Next, we consider the relic abundance of the lightest neutralino $\chi$, 
which is defined by
\begin{equation}
\chi=Z_{11} \tilde B + Z_{12} \tilde W_3 
+Z_{13} \tilde H_d^0 + Z_{14}\tilde H_u^0,
\end{equation}
where $Z_{1i}~(i=1\sim 4)$ are determined by 
diagonalizing the neutralino mass matrix
\begin{eqnarray}
M_N=\left(
\begin{array}{cccc}
M_{\tilde{B}}&0&-\cos\beta\sin\theta_Wm_Z&\sin\beta\sin\theta_Wm_Z\\
0&M_{\tilde{W}}&\cos\beta\cos\theta_Wm_Z&-\sin\beta\cos\theta_Wm_Z\\
-\cos\beta\sin\theta_Wm_Z&\cos\beta\cos\theta_Wm_Z&0&-\mu_H\\
\sin\beta\sin\theta_Wm_Z&-\sin\beta\cos\theta_Wm_Z&-\mu_H&0
\end{array}
\right).
\label{neutralino mass}
\end{eqnarray}
The annihilation of $\chi$ occurs through various
processes depending on its composition $Z_{1i}$ 
in the same way as the MSSM. Final states of the $\chi$ annihilation are
composed of all the SM particles which are lighter than $\chi$. 
The favorable parameter regions to explain the DM abundance by $\chi$ 
have been studied in detail \cite{susydm}.
We follow such studies in the present case keeping in mind 
the consistency with the parameters used in the neutrino sector.

The relic abundance of $\chi$ is determined by the mass and 
the composition $Z_{1i}$ which fixes the interaction of $\chi$ with the SM
particles. The parameters relevant to them are $\tan\beta$, $\mu_H$, 
soft supersymmetry breaking 
parameters $m_0$, $A$, $B$ and the gaugino masses $M_{\tilde W}$, 
$M_{\tilde B}$. Here we note that some of these parameters are related
each other.  $M_{\tilde W}=2M_{\tilde B}$ is expected at the weak scale
from the unification relation among gaugino masses.
Since $B$ is required to satisfy an electroweak symmetry breaking condition
$B=(m_{H_u}^2+m_{H_d}^2+2\mu_H^2)\sin 2\beta/2\mu_H$,
it is determined by $\mu_H $ if we fix the values of $m_{H_u}$,
$m_{H_d}$ and $\tan\beta$. In this analysis $m_{H_u}$ and $m_{H_d}$
are fixed to 500 GeV. We also take $\tan\beta=10$ which can be 
consistent with the discussion on the neutrino sector 
by tuning the value of $\lambda_u\lambda_d$.
Since other parameters are fixed to proper values at the weak scale,
$\mu_H$ and $M_{\tilde B}$ are treated as free parameters. 
Numerical calculation is executed by using the public code \texttt{micrOMEGAs} 
\cite{micromegas}.

\begin{figure}[t]
\begin{center}
\epsfxsize=8.5cm
\leavevmode
\epsfbox{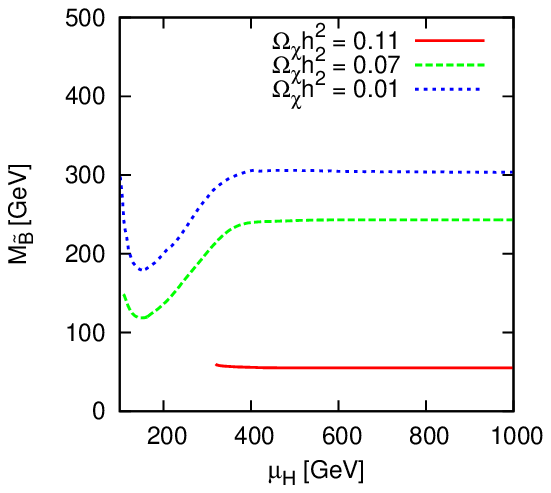}

\vspace*{-3mm}
{\footnotesize {\bf Fig.~4}~~
Contours of the relic abundance $\Omega_\chi h^2$ 
in the $(\mu_H, M_{\tilde B})$ plane. Only a region with
 $\mu_H>500$~GeV is allowed for $m_0=400$~GeV and $B>200$~GeV.}
\end{center}
\end{figure}

In Fig.~3, we plot contours of the mass $m_\chi$ and the composition
$Z_{1i}$ of $\chi$ in the $(\mu_H, M_{\tilde B})$ plane.
When we see this figure, we have to remind that the allowed region
should satisfy a condition $m_\chi<m_0$, which is required since $\chi$ is DM. 
Here we consider a case with $m_0=400$~GeV and $B>200$~GeV which are 
used in the study of neutrino sector for $M_1=3200$~GeV.
In this case $\mu_H>500$~GeV is required by the electroweak symmetry
breaking condition given above. Thus, Fig.~3 shows that $\chi$ is bino dominated
at the region with $M_{\tilde B}<400$~GeV where $\chi$ can be the
lightest superparticle.\footnote{In the $M_{\tilde B}>400$~GeV region, DM is a
sneutrino which is difficult to realize the right relic
abundance because of its effeicient annihilation.}
In Fig.~4, the contours of the relic abundance $\Omega_\chi h^2$ 
of $\chi$ is plotted in the $(\mu_H, M_{\tilde B})$ plane.
As found from Figs.~3 and 4, $\chi$ can be a DM component with the substantial
abundance at the above mentioned bino dominated region.
Since the $\psi_{N_1}$ abundance is not sensitive to the value 
of $m_0$ if $m_0<\mu_\eta$ is satisfied, we can find parameters for 
which $\chi$ is a DM component with substantial abundance 
under the condition (\ref{wmap}).
It is useful to note that in this model 
$\chi$ could be an important DM component 
in the region where it is rejected as the DM in the MSSM framework. 

\subsection{Decay of the right-handed neutrino dark matter}
In the previous part we showed that the DM can be composed of 
two components $\psi_{N_1}$ and $\chi$ which have the same order
abundance.
However, $\psi_{N_1}$ is not stable since
the $Z_2$ symmetry which guarantees its stability is not exact.
Since this symmetry is considered to be a remnant symmetry left 
after the spontaneous 
breaking of the anomalous U(1) at a high energy region, it is broken by
the anomaly effect. In fact,
the Green-Schwarz anomaly cancellation mechanism induces
the $Z_2$ violating interaction as the last term of $W$ nonperturbatively.
If $\psi_{N_1}$ is heavier than $\chi$, this interaction brings 
the decay of $\psi_{N_1}$ to $\chi$ through the diagrams shown in Fig.~5.

In order to examine whether $\psi_{N_1}$ can be dealt as the DM, we estimate 
the lifetime of $\psi_{N_1}$ due to the decay caused by 
this interaction.
It can be roughly estimated as 
  \begin{eqnarray}
 \tau_{\psi_{N_1}} \sim \left(\frac{ \mbox{3.2~TeV}}{M_1}\right)
 \left( \frac{\mu_\eta}{\mbox{3.6~TeV}} \right)^4
 \left( \frac{\mbox{0.25~TeV}}{B}\right)^2
\left( \frac{e^{2b_i}}{10^{77}}\right)\times 10^{26}  ~~\mbox{sec},
\label{life}
 \end{eqnarray}
where we use $|h_1|,~c_i\sim 1$ and $M_1 \gg m_0$. From this formula, 
we find that $\psi_{N_1}$ can have a 
sufficiently long lifetime compared with the age of the universe,
as long as $b_i > 79$ is satisfied.
Thus, although the true stable DM is the lightest neutralino $\chi$, 
we need to take account of the contribution of $\psi_{N_1}$ to the relic 
DM abundance and investigate the DM phenomenology.

Charged particle observation in the cosmic rays by PAMELA \cite{pamela}
and Fermi-LAT \cite{fermi}
suggests that there are deviations from the expected background.
The possibility has been discussed that these are consequences 
of the DM physics.
However, if we consider that they are yielded by the annihilation 
of the DM, the annihilation cross section required for the explanation of the 
relic abundance is too small \cite{mindep}.
Some enhancement of the annihilation cross section at the present
universe seems to be necessary \cite{sommerfeld,bwenhance}.   
On the other hand, if we consider the decay of the DM, these anomalies
found in the cosmic rays can
be understood as long as its lifetime is sufficiently long \cite{it,decay}.

\begin{figure}[t]
\begin{center}
\epsfxsize=4cm
\leavevmode
\epsfbox{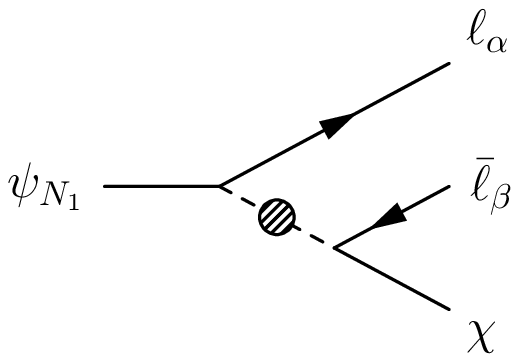}
\hspace*{7mm}
\epsfxsize=4cm
\leavevmode
\epsfbox{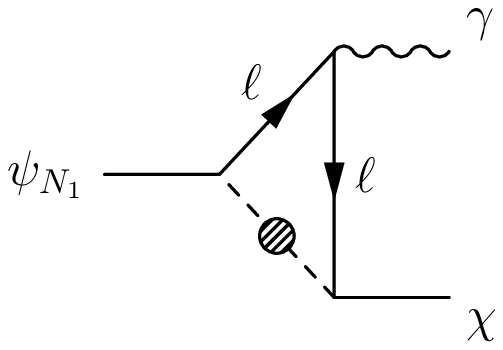}
\hspace*{7mm}
\epsfxsize=4cm
\leavevmode
\epsfbox{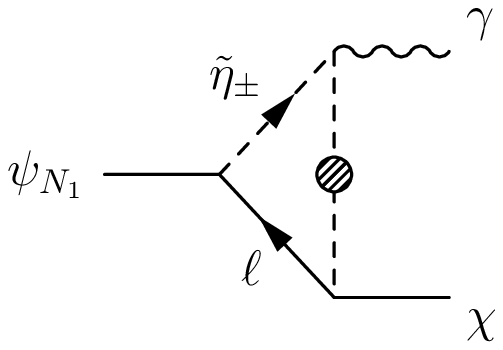}
\end{center}

\vspace*{-3mm}
{\footnotesize {\bf Fig.~5}~~Decay processes of $\psi_{N_1}$ to $\chi$.
A bulb represents the anomaly induced interaction $c_iBM_{\rm
 pl}e^{-b_i}\tilde L_i\tilde\eta_u$. }
\end{figure}

In the present model, particles yielded in the decay of 
$\psi_{N_1}$ may bring the required extra contributions to the cosmic rays. 
The expected flux depends on the scale of the $Z_2$ breaking 
$c_iBM_{\rm pl}e^{-b_i}$ in eq.~(\ref{softsb}) \cite{fks,anomu}.
In fact, if $b_i\sim 88$ is satisfied, the anomaly induced interaction
causes a large enhancement factor 
of $O(10^{77})$ in eq.~(\ref{life})
to realize a long lifetime of $O(10^{26})$ sec for 
$\psi_{N_1}$.
This lifetime is known to be suitable to explain the anomalies found in
the charged cosmic rays discussed above.
Moreover, since $\psi_{N_1}$ couples only with leptons 
and sleptons, this decay could yield only leptons and photon other than $\chi$. 
The flavor structure of neutrino Yukawa couplings (\ref{yukawa}) can restrict 
the final charged leptons to $\mu$ and $\tau$ dominantly.\footnote{Since
$\chi$ is dominated by the bino in our considering parameter region,
we need to impose $c_e=0$ in the anomaly induced interaction 
$c_iBM_{\rm pl}e^{-b_i}L_i\eta_u$ additionally in order to keep this feature.}
This feature makes the model favorable for the explanation of
the above mentioned cosmic ray anomalies.
In the following parts, we study the positron and electron flux predicted
by this decay process and compare it with the data. 
We also estimate the nature of photon flux expected in the radiative 
decay of $\psi_{N_1}$ which is shown in Fig.~5.

The metastable $\psi_{N_1}$ decays to $\chi\ell_\alpha\bar\ell_\beta$ 
through the left-handed diagram in Fig.~5. 
This decay is caused by the anomaly induced 
$\tilde\ell_\alpha$-$\tilde\eta$ mixing.
Since the mass of $\psi_{N_1}$ is of $O(1)$ TeV, 
$M_{\tilde \ell}(\simeq m_0)<M_1$ is 
naturally expected from a view point
of low energy supersymmetry. 
In this case
the intermediate slepton $\tilde\ell_\alpha$ is considered to be 
produced as an on-shell state. To take account of this possibility,
we use the propagator for the slepton $\tilde\ell_\alpha$ which 
contains the effect of decay width $\Gamma_{\tilde\ell_\alpha}$.

The differential decay width to the final state $\ell_\alpha$ is given by 
\begin{equation}
\frac{d \Gamma}{d E_\alpha}(\psi_{N_1}\to\ell_\alpha\bar{\ell}_\beta\chi)=
\frac{1}{4(4\pi)^3M_1}\int^\pi_0d\theta
F(\theta)\frac{2E_\alpha
E_\beta}{M_1+E_\alpha(\cos\theta-1)}\overline{|\mathcal{M}_{\alpha\beta}|^2},
\label{difwidth}
\end{equation}
where the $\theta$ is the angle between $\ell_\alpha$ and
$\bar{\ell}_\beta$. This formula is symmetric for the exchange of
$\ell_\alpha$ and $\bar\ell_\beta$.
The spin averaged amplitude $\overline{|\mathcal{M}_{\alpha\beta}|^2}$ 
is expressed as
\begin{eqnarray}
\overline{|\mathcal{M}_{\alpha\beta}|^2}&=&
\left(|A_\alpha|^2+|B_{\alpha\beta}|^2\right)
\left(c_\beta BM_{\mathrm{pl}}e^{-b}\right)^2
\frac{M_1E_\alpha\left(t-m_\chi^2\right)}
{(t-M_{\tilde{\ell}}^2)^2+M_{\tilde{\ell}}^2\Gamma_{\tilde{\ell}}^2}
\left(
\frac{\cos^2\theta_{\eta}}{t-m_{\eta_+}^2}+
\frac{\sin^2\theta_{\eta}}{t-m_{\eta_-}^2}\right)^2\nonumber\\
&+&
\left(|A_\beta|^2+|B_{\beta\alpha}|^2\right)
\left(c_\alpha BM_{\mathrm{pl}}e^{-b}\right)^2
\frac{M_1E_\beta\left(u-m_\chi^2\right)}
{(u-M_{\tilde{\ell}}^2)^2+M_{\tilde{\ell}}^2\Gamma_{\tilde{\ell}}^2}
\left(
\frac{\cos^2\theta_{\eta}}{u-m_{\eta_+}^2}+
\frac{\sin^2\theta_{\eta}}{u-m_{\eta_-}^2}\right)^2 \nonumber \\
&+&
|A_\alpha A_\beta|\left(c_\alpha c_\beta B^2M_{\mathrm{pl}}^2e^{-2b}\right)
\frac{M_1m_\chi\left(M_1^2+m_\chi^2-2M_1E_\chi\right)}
{\left[(t-M_{\tilde{\ell}}^2)^2+M_{\tilde{\ell}}^2
\Gamma_{\tilde{\ell}}^2\right]
 \left[(u-M_{\tilde{\ell}}^2)^2+M_{\tilde{\ell}}^2
\Gamma_{\tilde{\ell}}^2\right]}\nonumber\\
&\times&
\left(\frac{\cos^2\theta_\eta}{t-m_{\eta_+}^2}
+\frac{\sin^2\theta_\eta}{t-m_{\eta_-}^2}\right)
\left(\frac{\cos^2\theta_\eta}{u-m_{\eta_+}^2}
+\frac{\sin^2\theta_\eta}{u-m_{\eta_-}^2}\right).
\end{eqnarray}
In these formulas we use the definitions such as
\begin{eqnarray}
&&F(\theta)\equiv \sin\theta+(\pi-\theta)\cos\theta, \nonumber\\
&&E_\beta=\frac{M_1^2-m_\chi^2-2M_1E_\alpha}{2\left[M_1+
E_\alpha(\cos\theta-1)\right]},\qquad
E_\chi=\sqrt{E_\alpha^2+E_\beta^2+2E_\alpha E_\beta\cos\theta
+m_\chi^2},\nonumber \\
&&t\equiv M_1^2-2M_1E_\alpha,\qquad u\equiv M_1^2-2M_1E_\beta \nonumber \\
&&A_\alpha\equiv \frac{h_{\alpha
 1}^*}{\sqrt{2}}\left(g'Z_{11}+g Z_{12}\right),\qquad
 B_{\alpha\beta}\equiv h_{\alpha 1}^*h^E_\beta Z_{13},
\end{eqnarray}
where $g'$ and $g$ are the gauge coupling
constants for U(1)$_Y$ and SU(2)$_L$, respectively.
The mixing angle $\theta_\eta$ between $\eta_u$ and $\eta_d^\dag$,
can be taken as $\theta_\eta=\pi/4$ since their soft scalar masses 
are assumed to be universal.\footnote{
Even if they are not universal, however, this is a good approximation 
as long as $\mu_\eta$ is much larger than their soft scalar masses.}
In this derivation we use the universality for the slepton masses 
and $b_i$. We also assume
the flavor independent slepton decay width $\Gamma_{\tilde\ell}$.
Since we use the flavor structure (\ref{yukawa}) for neutrino 
Yukawa couplings and $c_e=0$,  suffices $\alpha$ and 
$\beta$ in eq.~(\ref{difwidth}) run over 
the lepton flavor $\mu$ and $\tau$.  
Thus, the decay of $\psi_{N_1}$ does not yield positron directly in the
final state.
 
In this model the positron is generated through the decay of 
$\mu^+$ and $\tau^\pm$. 
In the following positron flux calculation, we use the positron spectrum 
$\frac{dN_{\ell_\alpha e^+}}{dE}$ obtained from the simulation
by using the MONTE CARLO code in the public 
package \texttt{PYTHIA} \cite{code} which can generate 
the lepton $\ell_\alpha$ whose energy distribution 
is given by eq.~(\ref{difwidth}) and calculate the positron from the
decay of this lepton. 
The positron spectrum obtained through this calculation is shown in Fig.~6.
Although two leptons are contained in the final state of the
$\psi_{N_1}$ decay , the final positron flux is estimated by summing
the contribution from each lepton which can be treated independently
based on eq.~(\ref{difwidth}).
Thus, using this spectrum, the positron flux yielded through the decay 
of $\psi_{N_1}$ is expected to be observed at the earth as \cite{it}
\begin{eqnarray}
\Phi_{e^+}^{\rm prim}(E) &=&\frac{c}{4\pi M_1\tau_{\psi_{N_1}}}
\int_E^{E_{\rm max}} d E' G_{e^+}(E,E')~ \sum_{\alpha=e^+,\mu^+,\tau^{\pm}}
\mathrm{Br}_{\ell_\alpha}\frac{d N_{\ell_\alpha e^+}(E')}{d E'},
 \label{fe}
\end{eqnarray}
where $E_{\rm max} = (M_1^2-m_{\chi}^2)/2M_1$ and
$\mathrm{Br}_{\ell_\alpha}=
\sum_{\bar\ell_\beta}\Gamma(\psi_{N_1}
\rightarrow\ell_\alpha\bar\ell_\beta\chi)/\Gamma_{\rm tot}$
where $\Gamma_{\rm tot}$ is the total decay width of $\psi_{N_1}$, which
is fixed by including the final states with neutrinos.
This $\mathrm{Br}_{\ell_\alpha}$ is almost determined 
by $h^N_{\alpha 1}$ and $c_\alpha$
since we assume that the slepton masses and $b_\alpha$ in the last term
of the superpotential $W$ are universal. 
Moreover, $\mathrm{Br}_{\ell_\alpha}$ is determined by
$c_\alpha$ only in the present case since the flavor structure 
(\ref{yukawa}) for $h_{\alpha 1}^N$ is adopted. 
Values of $c_\alpha$ and $\mathrm{Br}_{\ell_\alpha}$ 
used here are shown in Table 2. 

\begin{table}[b]
\begin{center}
\begin{tabular}{|c|c|c|c|}
\hline
&case (a)& case (b)&case (c)\\ \hline
$c_e:c_\mu:c_\tau$ & 0 : 1 : 1 & 0 : 1 : 0 & 0 : 0 : 1\\ \hline
$\mathrm{Br_{e^+}}$ : $\mathrm{Br_{\mu^+}}$ :
 $\mathrm{Br_{\tau^+}}$
 : $\mathrm{Br_{\tau^-}}$
&0 : $\frac{1}{4}$ :$\frac{1}{4}$  : $\frac{1}{4}$  
& 0 : $\frac{5}{12}$ : $\frac{1}{12}$ : $\frac{1}{12}$
& 0 : $\frac{1}{12}$ : $\frac{5}{12}$ : $\frac{5}{12}$ \\ \hline
\end{tabular}
\end{center}
\vspace*{3mm}

{\footnotesize{\bf Table 2}~~ Values of $c_\alpha$ and
 $\mathrm{Br}_{\ell_\alpha}$ used in the calculation for the positron. 
It should be noted that $\mathrm{Br}_{\mu^+}$ ($\mathrm{Br}_{\tau^\pm}$) 
takes a nonzero value even if $c_\mu$ ($c_\tau$) is zero.}
\end{table}

The positron Green's function $G_{e^+}$ can be approximately written 
as \cite{it,Hisano:2005ec}
\begin{eqnarray}
 G_{e^+}(E,E') &\simeq
  &\left(\frac{\Omega_{\psi_{N_1}}}{\Omega_{\psi_{N_1}}+\Omega_\chi}\right)
 \frac{10^{16}}{E^2} 
 \exp [ a+b(E^{\delta-1}-E^{\prime \delta -1})   ]~~\mbox{cm}^{-3}~\mbox{s},
 \end{eqnarray}
 where $a,b$ and $\delta$ depend on the diffusion model 
and the assumed halo profile \cite{it,Moskalenko:1997gh,
med}. 
Since the result is known not to be heavily dependent on these in the
case of DM decay, we use the MED model \cite{med} and the NFW 
profile \cite{Navarro:1995iw}.
They fix these parameters to $a=-1.0203, b=-1.4493$ and $\delta=0.70$.
We also assume that the two DM components $\psi_{N_1}$ and $\chi$ 
have the same density profile in our galaxy.

\begin{figure}[t]
\begin{center}
\epsfxsize=8cm
\leavevmode
\epsfbox{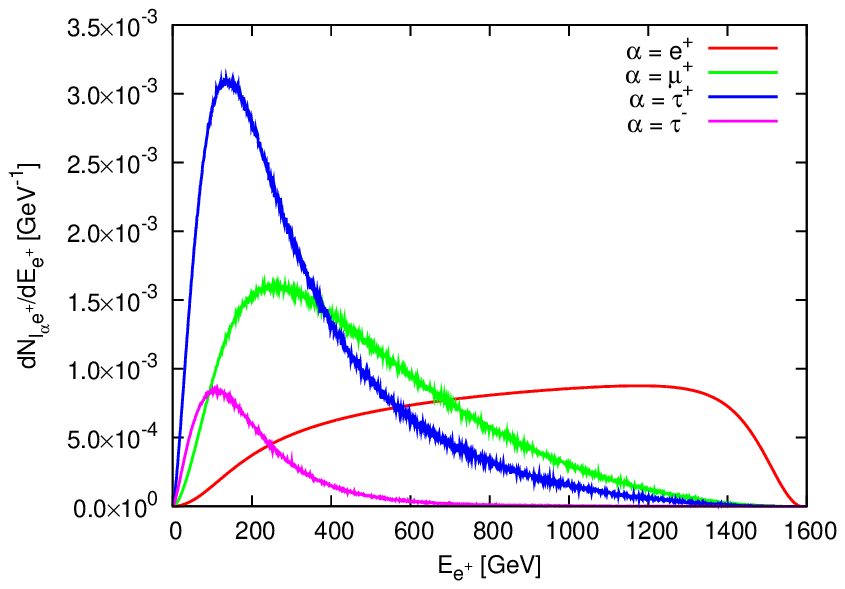}
\end{center}
\vspace*{-3mm}

{\footnotesize {\bf Fig.~6}~~The positron spectrum $\frac{dN_{\ell_\alpha
 e^+}}{dE}$ calculated from $\frac{d\Gamma}{dE_\alpha}$ by using
 \texttt{PYTHIA}. $M_1=3.2~\mathrm{TeV}$ and $m_\chi=300~\mathrm{GeV}$ 
are taken as a typical example here.}
\end{figure}

The background flux for electrons and positrons in the cosmic rays 
are given by \cite{postronf}
\begin{eqnarray}
&&\Phi^{\rm prim. bkg}_{e^-}(E) =N_{\phi}\frac{0.16 E^{-1.1}}{
1+11 E^{0.9}+3.2 E^{2.15}},
\nonumber \\
&&\Phi^{\rm sec. bkg}_{e^-}(E) =N_{\phi}\frac{0.7 E^{0.7}}{
1+11 0E^{1.5}+600 E^{2.9}+580 E^{4.2}},
\nonumber \\
&&\Phi^{\rm sec. bkg}_{e^+}(E) =N_{\phi}\frac{4.5 E^{0.7}}{
1+650 E^{2.3}+1500 E^{4.2}}
\label{backg}
\end{eqnarray}
in the unit of $[\mbox{GeV}\cdot\mbox{cm}^2\cdot\mbox{sec}\cdot\mbox{sr}]^{-1}$.
The energy $E$ is in the GeV unit, and the normalization 
factor $N_{\phi}$ is fixed to be $N_\phi=0.66$ in the present analysis.
By using these results, we estimate the quantities reported by 
PAMELA and Fermi-LAT. They are expressed as
\begin{eqnarray}
&&\frac{\Phi_{e^+}}{\Phi_{e^+}+\Phi_{e^-}} =
\frac{\Phi^{\rm prim}_{e^+}+
\Phi^{\rm sec. bkg}_{e^+} }{\Phi^{\rm prim}_{e^+}+
\Phi^{\rm sec. bkg}_{e^+}+\Phi^{\rm prim}_{e^-}+\Phi^{\rm prim.bkg}_{e^-}
+\Phi^{\rm sec. bkg}_{e^-}} \qquad\mbox{for PAMELA},  \nonumber \\
&&\Phi_{e^+}+\Phi_{e^-}= \Phi^{\rm prim}_{e^+}+
\Phi^{\rm sec. bkg}_{e^+}+\Phi^{\rm prim}_{e^-}+\Phi^{\rm prim.bkg}_{e^-}
+\Phi^{\rm sec. bkg}_{e^-}  
\qquad\mbox{for Fermi-LAT}, \nonumber
\end{eqnarray}
respectively.
In this estimation we should use the parameters which satisfy the
constraints from the neutrino oscillation data, the LFV and the WMAP
data. Such examples are shown in Figs.~1 and 2.
Although the ambiguity exists in the choice of each value of these
parameters, it is absorbed into the assumed $\psi_{N_1}$ lifetime
$\tau_{\psi_{N_1}}$. It can be justified by tuning the free parameter $b_i$
as found from eq.~(\ref{life}).\footnote{It is interesting that the
required value for $b_i$ can be consistent with the ones which explain
the hierarchy of the coupling constants and the masses \cite{anomu}.} 
Thus, the lifetime $\tau_{\psi_{N_1}}$ is treated as a free parameter 
in this analysis.

The fluxes predicted by the model are plotted in each frame of Fig.~7
for some typical values of $\tau_{\psi_{N_1}}$ and $c_\alpha$ 
listed in Table 2.
The data of PAMELA \cite{pamela} and Fermi-LAT \cite{fermi} are
also plotted in the corresponding frame. 
We can fit the predicted flux in all the cases shown in Table 2 to 
the data of PAMELA well. However, the situation is different in the
Fermi-LAT case. Although the predicted flux in the case (a) and (b) can
be fitted to the observed data well, the case (c) can not be fitted
to the data. The reason is that the positron produced 
from $\tau^\pm$ is softer than the one from $\mu^+$.
We find that the best fit is obtained in the case (b) where $\psi_{N_1}$
decays to $\mu^+$ dominantly. This case is also allowed 
from a view point of the constraint of diffuse gamma ray \cite{mpsv}. 
Since we aim to explain both anomalies in this analysis, we need to
suppose a large mass for $\psi_{N_1}$. It requires large neutrino Yukawa
couplings as seen in the previous part. However, if we confine our study
to explain the PAMELA anomaly only, rather light $\psi_{N_1}$ can also 
work well. In that case neutrino Yukawa couplings need not to be so large. 
 
\begin{figure}[t]
\begin{center}
\epsfxsize=7.5cm
\leavevmode
\epsfbox{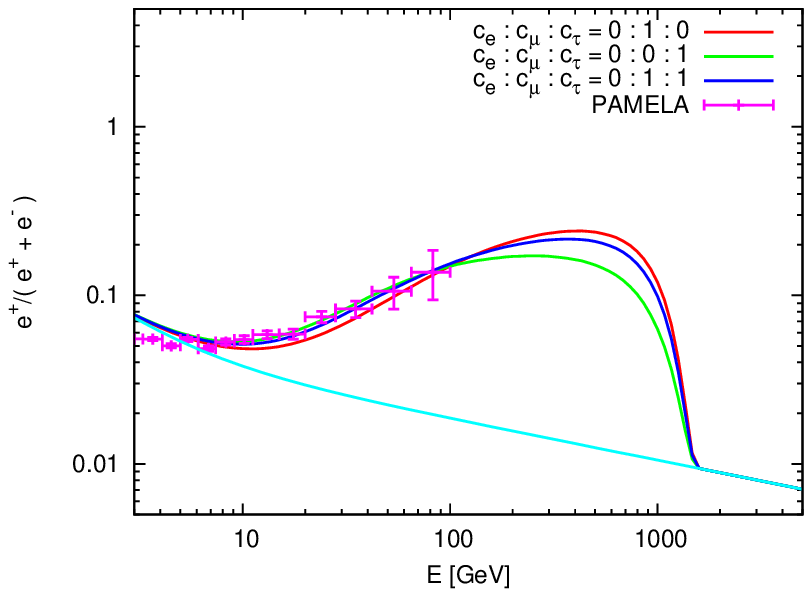}
\hspace{7mm}
\epsfxsize=7.5cm
\leavevmode
\epsfbox{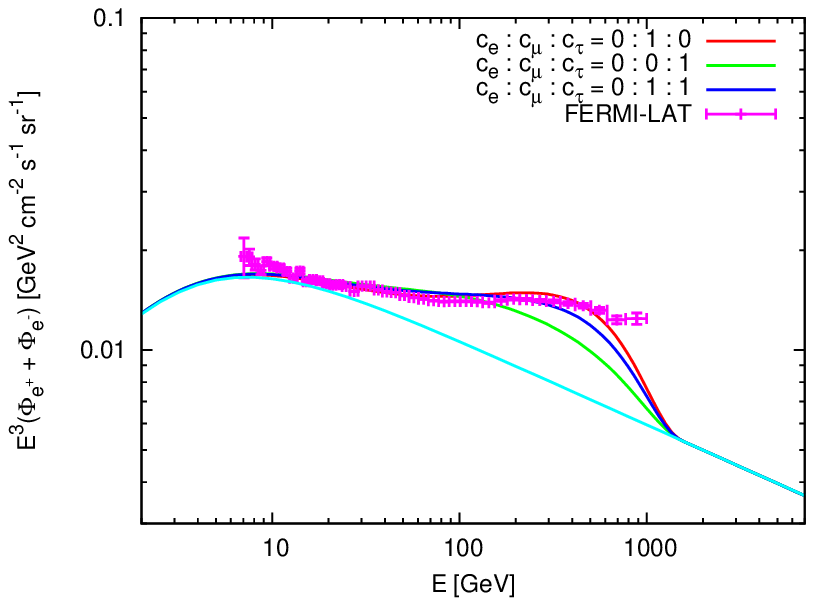}
\end{center}
{\footnotesize {\bf Fig.~7}~~Flux of positron and electron predicted 
by the model for the data of PAMELA 
(the left frame) and Fermi-LAT (the right frame). 
Relevant parameters are fixed as $M_1=3.2\:\mathrm{TeV}$ and 
$m_\chi=300\:\mathrm{GeV}$. The flux is plotted for three typical cases
 such as  $(c_e : c_\mu : c_\tau,~\tau_{\psi_{N_1}})=$ 
(0 : 1 : 1, $4.0\times 10^{26}$s), (0 : 1 : 0, $6.7\times 10^{25}$s) 
and (0 : 0 : 1, $3.3\times 10^{25}$s).}
\end{figure}
 
The heavier DM component $\psi_{N_1}$ has also a radiative decay mode to the
lightest neutralino $\chi$.
Its one-loop diagram is shown in Fig.~5.
This decay associates a characteristic gamma which can be
found through the observation of the cosmic gamma rays.
It has a line shape spectrum at the energy $(M_1^2-m_{\chi}^2)/2M_1$ 
which corresponds to the endpoint of the gamma ray spectrum 
generated through the processes such as the bremsstrahlung and 
the inverse Compton scattering associated to the $\psi_{N_1}$ decay and
also through the hadronization, fragmentation and decay of the final
states.

The width of this radiative decay is calculated as
\begin{equation}
\Gamma_\gamma=\frac{e^2}{8(4\pi)^5}\frac{\left(M_1^2-m_\chi^2\right)^3}{M_1^3}
\left(|\mathcal{A}|^2+|\mathcal{B}|^2\right),
\end{equation}
where ${\cal A}$ and $\mathcal{B}$ are defined as
\begin{eqnarray}
{\cal A}&=&{\cal G}M_1
\left(\cos^2\theta_\eta
 I(M_1^2,m_\chi^2,m_{\eta_+}^2,m_{\tilde{\ell}}^2)+
\sin^2\theta_\eta
I(M_1^2,m_\chi^2,m_{\eta_-}^2,m_{\tilde{\ell}}^2)\right) \nonumber\\
&&+{\cal G}^*m_\chi\left(\cos^2\theta_\eta
	   I(m_\chi^2,M_1^2,m_{\tilde{\ell}}^2,m_{\eta_+}^2)+\sin^2\theta_\eta
	   I(m_\chi^2,M_1^2,m_{\tilde{\ell}}^2,m_{\eta_-}^2)\right), \nonumber\\
\mathcal{B}&=&{\cal G}m_\chi
\left(\cos^2\theta_\eta
 I(m_\chi^2,M_1^2,m_{\tilde{\ell}}^2,m_{\eta_+}^2)+\sin^2\theta_\eta
 I(m_\chi^2,M_1^2,m_{\tilde{\ell}}^2,m_{\eta_-}^2)\right),\nonumber\\
&&+{\cal G}^*M_1\left(\cos^2\theta_\eta I(M_1^2,m_\chi^2,m_{\eta_+}^2,
m_{\tilde{\ell}}^2)+\sin^2\theta_\eta
	I(M_1^2,m_\chi^2,m_{\eta_-}^2,m_{\tilde{\ell}}^2)\right), \nonumber\\
{\cal G}&=&
\left(g'Z_{11}+gZ_{12}\right)\left(M_{\mathrm{pl}}B^*\right)
\left(\sum_\alpha h_{\alpha 1}^*c_{\alpha}^*e^{-b_\alpha}\right).
\end{eqnarray}
In these formulas we use the definitions such as\footnote{Although 
the function $I(m_a^2,m_b^2,m_c^2,m_d^2)$ may be considered singular at
$m_c^2=m_d^2$ for example, one can check that it is not singular.}
\begin{eqnarray}
I(m_a^2,m_b^2,m_c^2,m_d^2)&=&\frac{1}{2m_b^2(m_c^2-m_d^2)}\left[
I_1\left(\frac{m_a^2}{m_b^2},\frac{m_d^2}{m_b^2}\right)-I_1
\left(\frac{m_a^2}{m_b^2},\frac{m_c^2}{m_b^2}\right)
\right] \nonumber \\
&&+\frac{1}{m_a^2(m_c^2-m_d^2)}\left[
I_2\left(\frac{m_b^2}{m_a^2},\frac{m_d^2}{m_a^2},\frac{m_c^2}{m_a^2}\right)
-I_2\left(\frac{m_b^2}{m_a^2},\frac{m_d^2}{m_a^2},\frac{m_d^2}{m_a^2}\right)
\right]\nonumber \\
&&+\frac{1}{m_b^2(m_c^2-m_d^2)}\left[
I_3\left(\frac{m_a^2}{m_b^2},\frac{m_c^2}{m_b^2},\frac{m_d^2}{m_a^2}\right)
-I_3\left(\frac{m_a^2}{m_b^2},\frac{m_c^2}{m_b^2},\frac{m_c^2}{m_b^2}\right)
\right], \nonumber 
\end{eqnarray}
\begin{eqnarray}
&&I_1\left(\alpha_1,\alpha_2\right)=
\left(\frac{1-\alpha_2}{1-\alpha_1}\right)^2
\log\left|\frac{\alpha_1-\alpha_2}{1-\alpha_2}\right|
-\left(\frac{\alpha_2}{\alpha_1}\right)^2
\log\left|\frac{\alpha_1-\alpha_2}{\alpha_2}\right|
+\frac{\alpha_1-\alpha_2}{\alpha_1(1-\alpha_1)}, \nonumber \\
&&I_2\left(\alpha_1,\alpha_2,\alpha_3\right)=
\int_0^1dx\frac{x(1-x)}{x(1-\alpha_1)+(\alpha_2-\alpha_3)}
\left[1+\frac{x\alpha_1-\alpha_2}{x(1-\alpha_1)
+(\alpha_2-\alpha_3)}\log\left|\frac{\alpha_2-\alpha_1x}
{\alpha_3-x}\right|\right], \nonumber \\
&&I_3\left(\alpha_1,\alpha_2,\alpha_3\right)=
\int_0^1dx\frac{x(1-x)}{x(1-\alpha_1)+(\alpha_2-\alpha_3)}
\left[1+\frac{x-\alpha_3}{x(1-\alpha_1)+(\alpha_2-\alpha_3)}
\log\left|\frac{\alpha_2-\alpha_1x}{\alpha_3-x}\right|\right].\nonumber \\
\end{eqnarray}
The contribution from the Higgsino component can be neglected
since it is proportional to the lepton mass and then small enough. 

If we use these formulas, we can estimate the diffuse gamma flux 
generated by the $\psi_{N_1}$ decay. 
For example, we could predict the monochromatic gamma ray flux generated 
through this DM decay in the Milky Way halo as
\begin{eqnarray}
\Phi_{\rm halo}^\gamma&=&
\frac{\Gamma_\gamma \Omega_{\psi_{N_1}}}
{\Omega_{\psi_{N_1}}+\Omega_\chi}\frac{1}{4\pi
M_1}\int_{\ell.o.s}d\vec{\ell}\rho^{\rm MW}_{\psi_{N_1}}(\vec{\ell}),
\end{eqnarray} 
where an integral is done over the DM distribution along a line of
sight. All astrophysical uncertainty is contained in this integral.
The radiative decay width $\Gamma_\gamma(\psi_{N_1}\to
\chi\gamma)$ is plotted as a function of the mass of $\psi_{N_1}$ in Fig.~8.
In this calculation we adopt the case (b) in Table 2 and use the
following parameters:
\begin{eqnarray}
&& \bar\lambda=7.74\times 10^{-9}, \quad m_0=400~{\rm GeV}, \quad
 M_3=9000~{\rm GeV}, \quad \varphi_1-\varphi_2=\frac{\pi}{4}, \nonumber \\
&&\tan\beta=10, \quad \mu_\eta=3600~{\rm GeV}, \quad  B=250~{\rm GeV},
\quad M_{\tilde B}=300~{\rm GeV}.
\label{para1}
\end{eqnarray}
These are the same ones used in the estimation of both the relic 
abundance of $\psi_{N_1}$ and the positron flux generated 
by the $\psi_{N_1}$ decay.
This result suggests that it may be observed at the proposed Cherenkov
Telescope Array in the future \cite{gam}.
If this line shape gamma flux is observed,
we can consider that it is a signature of the model for its peculiarity. 

\begin{figure}[t]
\begin{center}
\epsfxsize=8cm
\leavevmode
\epsfbox{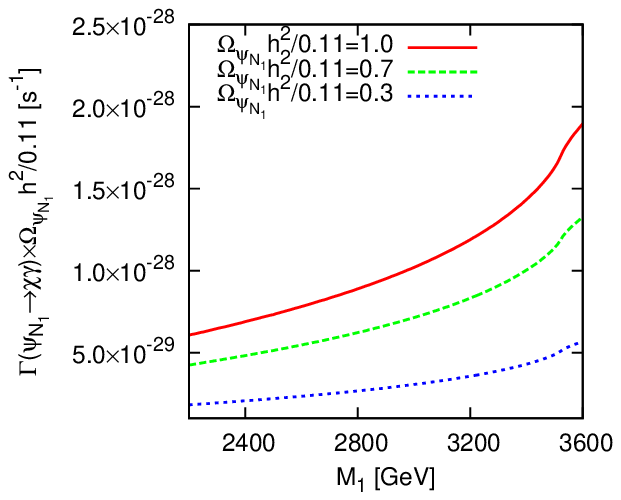}

\vspace*{-5mm}
{\footnotesize {\bf Fig.~8}~~The decay width 
for $\psi_{N_1}\to\chi\gamma$ as a function of the mass of $\psi_{N_1}$.}
\end{center}
\end{figure}

\subsection{Direct detection of the neutralino}
Direct detection of the DM can clarify the nature of DM \cite{susydm}.
Several experiments to search its elastic scattering with nuclei
such as CDMSII, XENON100 and XMASS are now under going or will 
start in near future. Some of these experiments have already constrain
the models. For example, a vast region of the parameter space in the
CMSSM has been excluded \cite{susydm}. 
Thus, it is crucial to address the discriminative features 
of the present model, which could be expected to be found through 
these experiments.  
In our model there are two components of the DM, $\psi_{N_1}$ and $\chi$.
Since $\psi_{N_1}$ does not have interactions with
nuclei at tree level, the scattering cross section is heavily
suppressed by the loop factor. Thus, it is difficult to detect 
it in these experiments.
On the other hand, the lightest neutralino $\chi$ can be scattered with nuclei 
at tree level since it has the same nature as the ordinary neutralinos 
in the MSSM. 
However, the constraint on the $\chi$ mass and its 
scattering cross section imposed by the relic abundance can be 
different from the one in the MSSM as discussed in the previous part. 
Since the model has two DM components, the relic abundance
constraint should be satisfied by both of these as shown in eq.~(\ref{wmap}).
Therefore, conditions for the parameters relevant to the direct
search of $\chi$ can be changed from the one in the MSSM, although
the interactions of $\chi$ with quarks are same as the MSSM neutralino.
This could give a new possibility for the direct search experiments,
which is not allowed in the MSSM case.

The spin independent scattering cross section between 
the neutralino $\chi$ and  
the nucleus with the atomic number $Z$ and the mass number $A$ 
is expressed as \cite{susydm}
\begin{equation}
\sigma_N^{\rm SI}=\frac{4 m_r^2}{\pi}\left[Zf_p + (A-Z)f_n\right]^2,
\end{equation} 
where no momentum transfer is assumed.
In case of large squark masses $m_\chi\ll m_{\tilde q}$, the effective
couplings of the neutralino $\chi$ with the proton ($f_p$) and 
the neutron ($f_n$) are written as
\begin{equation}
\frac{f_{p,n}}{m_{p,n}}\simeq \sum_{q=u,d,s}\frac{f_{T_q}^{p,n}f_q}{m_q}
+\frac{2}{27}f_{TG}^{p,n}\sum_{q=c,b,t}\frac{f_q}{m_q}. 
\end{equation}
where $f_q$ is the scalar four-point effective coupling constant whose
concrete expression can be found in \cite{susydm}.
$f_{T_q}$ represents the matrix element of nucleon defined 
by $\langle N|\bar qq|N\rangle=f_{T_q}M_n/m_q$  
and $f_{TG}$ is expressed as $\displaystyle f_{TG}=1-\sum_{q=u,d,s}f_{T_q}$.
Although we use the values of $f_{T_{u,d}}$ given in \cite{susydm},
we adopt the smaller value of $f_{T_{s}}$ which is given in \cite{ts}, 
\begin{eqnarray}     
&&f_{T_u}=0.023, \quad f_{T_d}=0.034, \quad f_{T_s}=0.02,  
\qquad {\rm for}~N=n, \nonumber \\
&&f_{T_u}=0.019, \quad f_{T_d}=0.041, \quad f_{T_s}=0.02,  
\qquad {\rm for}~N=p. 
\end{eqnarray}
In the numerical calculation, we treat $\mu_H$ and $M_{\tilde B}$ as 
free parameters in the allowed range shown in Figs.~3 and 4. 
Other relevant parameters are fixed to $\tan\beta=10$ and 
$M_{\tilde\ell}=m_\chi+50$~GeV where $M_{\tilde\ell}$ is the slepton
mass. Squark masses are assumed to be heavy enough. These parameters are
those used in Figs.~3 and 4.

\begin{figure}[t]
\begin{center}
\epsfxsize=7.5cm
\leavevmode
\epsfbox{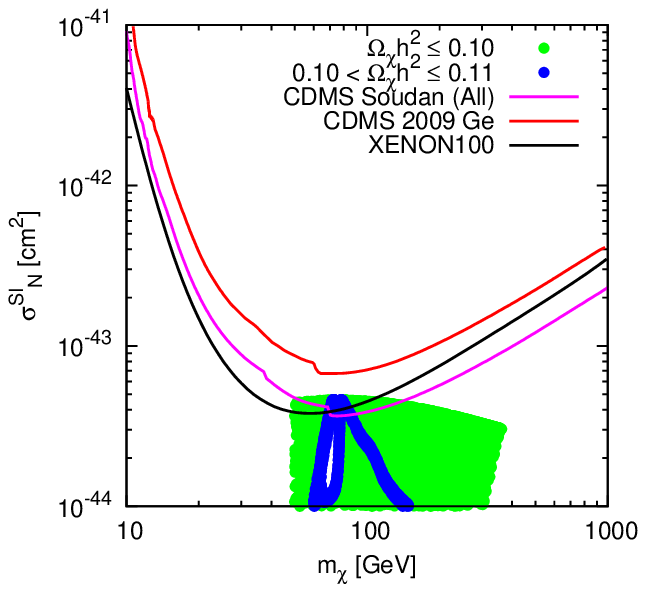}
\hspace*{3mm}
\epsfxsize=7.5cm
\leavevmode
\epsfbox{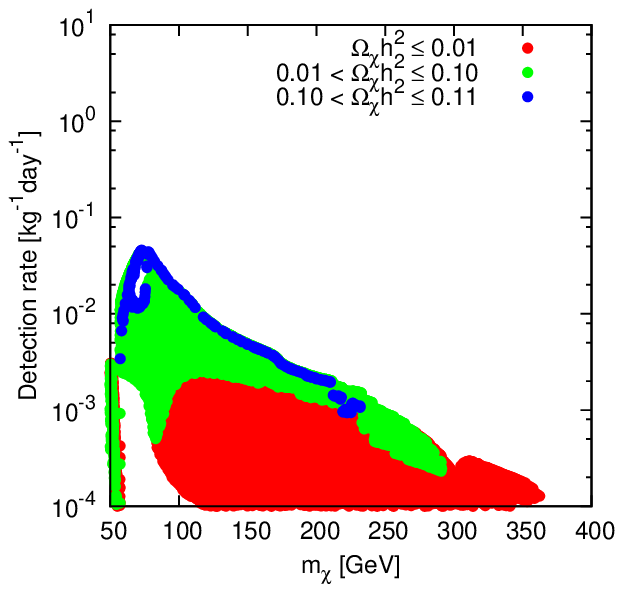}
\end{center}
\vspace*{-3mm}
{\footnotesize {\bf Fig.~9}~~The left frame shows the allowed region in the 
$(m_\chi, \sigma_N^{\mathrm{SI}})$ plane for the certain relic abundance
 of $\chi$. The right frame shows the detection rate of $\chi$ 
in the XENON target for the $\chi$ mass and the certain relic abundance
 of $\chi$. The MSSM corresponds to the upper edge of the region 
colored by blue.}
\end{figure}

We use \texttt{micrOMEGAs} \cite{micromegas} in the analysis of 
the spin independent cross section with nucleon and the detection 
rate of the $\chi$.\footnote{The code includes the contributions 
from the $\chi$-gluon interaction via heavy quark loops.}
The left frame of Fig.~9 shows the region in the plane of $m_\chi$ and
$\sigma^{\mathrm{SI}}_N$ which is predicted by the model for various
values of the relic abundance of $\chi$.
The blue region corresponds to $0.10< \Omega_\chi h^2\le 0.11$
which includes the MSSM. On the other hand, the green region stands for
the one with $\Omega_\chi h^2\le 0.10$ which can be consistent with the
WMAP data in this model.  
The CDMSII and XENON100 bounds are also plotted by a violet, red and black solid line 
in this frame respectively\cite{cdms}. 
The region consistent with the relic abundance required for
$\chi$, one of the DM components, is found to be much extended 
in comparison with the MSSM case. 
This occurs since the relic abundance constraint becomes much weaker than the
MSSM case such as $\Omega_\chi h^2 < \Omega_{\mathrm{MSSM}}h^2\simeq 0.11$.
The right frame shows the detection rate of $\chi$ expected 
in the XENON target for each value of $m_\chi$ and the relic 
abundance of $\chi$.
It suggests that the detection rate can be decreased by order one 
compared with the MSSM case as long as 
$\Omega_{\psi_{N_1}}h^2$ and $\Omega_\chi h^2$ are comparable.
These features could allow us to distinguish this model from the MSSM.

\section{Conclusion}
We have studied the nature of the DM sector in a supersymmetric
extension of the radiative neutrino mass model. 
An anomalous U(1) symmetry is introduced to explain the hierarchical 
structure of the coupling constants and mass scales in the model.
The spontaneous breaking of this symmetry can induce a new $Z_2$
symmetry which guarantees the stability of the lightest odd parity particle. 
As a result, the model has two DM components as long as $R$ parity is
assumed to be conserved. However, since one of these discrete symmetries 
which guarantee the stability of DM is not exact due to the anomaly, 
one DM component is unstable to decay through a hugely suppressed 
term which is nonperturbatively induced via the anomaly effect.
These DM components could be detected through the indirect search of the
yields of the decaying DM and the direct search of the elastic
scattering from nuclei by taking account that the DM relic abundance is
composed of these. Positrons generated by the decaying DM can explain
the cosmic ray anomaly reported recently. Parameter regions predicted
by the direct detection can be different from the MSSM case since two DM
components may contribute the relic abundance in the same order.
If the line shape gamma is observed in the cosmic ray, we might 
confirm the model by combining it with the direct search of the DM. 
Forth coming experiments for DM can give fruitful information to 
the model.

\vspace*{5mm}
We would like to thank Martin Holthausen for careful reading of the manuscript.
This work is partially supported by a Grant-in-Aid for Scientific
Research (C) from Japan Society for Promotion of Science (No.21540262)
and also a Grant-in-Aid for Scientific Research on Priority Areas 
from The Ministry of Education, Culture, Sports, Science and Technology 
(No.22011003).
The numerical calculations were carried out on SR16000 at YITP in
Kyoto University.
\vspace*{7mm}

\newpage
\bibliographystyle{unsrt}

\end{document}